\documentclass[twocolumn,aps,pra,showpacs,floatfix]{revtex4}

\usepackage{dcolumn}
\usepackage{graphicx}
\usepackage{epsf}
\usepackage{amsmath}
\usepackage{bm}
\usepackage{bbm}
\usepackage{color}

\newcommand{\dd}{\mathrm d}

\newcommand{\ii}{\mathrm i}
\newcommand{\naively}{na$\ddot{\mbox{\i}}$vely}
\newcommand{\naive}{na$\ddot{\mbox{\i}}$ve}

\begin{document}

\sloppy

\title{Virtual Resonant States in Two-Photon Decay Processes:\\
Lower--Order Terms, Subtractions, and Physical Interpretations}

\author{Ulrich D.~Jentschura}

\affiliation{Department of Physics, Missouri University of Science
and Technology, Rolla MO65409-0640, USA}
\affiliation{Max--Planck--Institut f\"ur Kernphysik,
Postfach 10 39 80, 69029 Heidelberg, Germany}

\begin{abstract}
We investigate the 
two-photon decay rate of a highly excited atomic state 
which can decay to bound states of lower energy 
via cascade processes.  We show that a \naive\  treatment of the process,
based on the introduction of phenomenological decay rates 
for the intermediate, resonant states, leads to lower-order terms
which need to be subtracted in order to obtain the coherent
two-photon correction to the decay rate. The sum of the lower-order 
terms is exactly equal to the one-photon decay rate of the initial
state, provided the \naive\ two-photon decay rates
are summed over all available two-photon channels. 
A quantum electrodynamics (QED) treatment of the problem leads to 
an ``automatic'' subtraction of the lower-order terms.
\end{abstract}

\pacs{12.20.-m,12.20.Ds,31.30.jc}

\maketitle

%
% Introduction
%
\section{Introduction}

The problem of the calculation of the two-photon decay rate from the hydrogen 
$2S$ to the $1S$ state was solved in 1931 by Maria 
G\"{o}ppert--Mayer~\cite{GM1931}. Essentially, G\"{o}ppert--Mayer interpreted
the two-photon decay in second-order time-dependent
perturbation theory as a combined transition,
from $2S$ to a virtual $P$ state and then, in a second step, to
the ground state. However, if one generalizes the 
problem trivially, namely to the $3S \to 1S$ two-photon 
decay, then one inevitably encounters
calculational problems due to the presence of a resonant, intermediate,
virtual $2P$ state which leads to a quadratic singularity 
along the photon energy integration contour (see Refs.~\cite{Fl1984,%
TuYeSaCh1984,CrTaSaCh1986,FlPaSt1987,FlScMi1988,ChSu2007}).
Formally, thus, in the limit of vanishing decay width
of the intermediate resonant $2P$ state, the total two-photon 
decay rate for $3S \to 1S$  would be infinitely large, provided we
apply the formalism employed by G\"{o}ppert--Mayer to the 
$3S \to 1S$ decay without any modifications or
regularizations. 

Of course, this problem has been realized, 
and three possible considerations have been 
used in order to circumvent it:
(i) We may observe that the $3S$
state can decay via one-photon, electric-dipole decay
($3S \to 2P$, see p.~266 of Ref.~\cite{BeSa1957}),
and we can therefore safely assume that the 
decay rate of $3S$ will be given by the one-photon decay rate
$3S \to 2P$ to an excellent approximation, 
and thus we may entirely neglect the two-photon contribution.
(ii) Realizing that the virtual $2P$ state causes problems,
one can speculate about a possible exclusion of this 
state from the sum over the virtual, intermediate 
states in the propagators~\cite{Fl1984,TuYeSaCh1984,%
CrTaSaCh1986,FlPaSt1987,FlScMi1988}.
However, it has been argued in Ref.~\cite{Je2008}
that this procedure is not gauge invariant and therefore
cannot lead to a consistent solution of the problem.
(iii) Recently~\cite{Je2007,JeSu2008}, it has been pointed out that the 
divergences associated with the quadratic singularities 
do not occur if one interprets the 
two-photon decay rate as the imaginary part
of the two-loop self-energy. This observation is in full analogy to the 
one-loop self-energy whose imaginary part gives the 
one-photon decay rate~\cite{BaSu1978}. In the treatment used in
Refs.~\cite{Je2007,Je2008,JeSu2008}, one obtains
expressions for the two-photon decay rate which are
finite as the regulators (infinitesimal $\ii \epsilon$ terms in the 
propagator denominators) approach zero at the end of the 
calculation, even if we investigate the problematic
decay modes $3S \to 1S$, $4S \to 1S$ etc.
Besides the analogy with the one-photon decay rate/one-photon
self-energy relations, 
the treatment used in Refs.~\cite{Je2007,Je2008,JeSu2008} 
has been made plausible by field-theoretical arguments
and analogies to radiative corrections to cascade
processes in photon emission in crossed electric-magnetic 
fields~\cite{Ri1972} and by analogies with the 
treatment of quadratic singularities in Lamb-shift calculations 
for highly excited states~\cite{JeSoMo1997}.

However, none of these treatments answer the simple
question: What do we do about the expressions used by Maria
G\"{o}ppert--Mayer, when generalized to the $3S \to 1S$ or $4S \to 1S$
decays in an obvious way? What if we regularize the divergences due to 
virtual low-lying $P$ states by the most physical, most obvious 
regularization available, namely the total, physical 
decay rate of those intermediate, resonant, $P$ states
that we insert into the propagator denominators?
(This approach has been discussed at various places in the 
literature, e.g., p.~2447 of Ref.~\cite{Fl1984} or 
p.~4 of Ref.~\cite{ChSu2007}.)
How can we interpret the resulting expressions from this
perhaps \naive{}, but physically intuitive and plausible approach?

We attempt to answer these questions here.
Essentially, we find that the \naive{} generalization of the 
expressions used by G\"{o}ppert--Mayer to the $3S \to 1S$ and 
$4S \to 1S$ decays is perfectly reasonable, provided we
subtract a lower-order term which is hidden in the 
expressions for the $3S \to 1S$ and $4S \to 1S$ decays,
but absent in the $2S \to 1S$ decays. After the subtraction,
the result obtained using the \naive{} formalism
is in perfect agreement with the result obtained 
for the two-photon decay rate from the 
imaginary part of the two-photon self-energy. 
We reemphasize that the result for the $2S \to 1S$ decay rate
obtained by G\"{o}ppert--Mayer does not require any additional 
subtractions or modifications. However,
one should realize that in order to extract physically valid information
from the \naive{} expressions for the 
two-photon decay rates from highly excited states, like
$3S \to 1S$ and/or $4S \to 1S$ etc.,
one first has to subtract a lower-order term, whose physical
interpretation is also discussed in the current paper.

The general outline of this paper is as follows.
In Sec.~\ref{gen}, we qualitatively explain the essence
of the considerations needed to fully understand the 
removal of the lower-order terms which are present in the 
\naive\  expression for the two-photon decay rate
and thereby anticipate a few considerations to be explained in 
greater detail in the following.  In Sec.~\ref{go}, we then calculate the 
lower-order terms explicitly for the case of 
the $4S \to nS$ decays ($n = 1,2,3$), and explain their removal.
Conclusions are reserved for Sec.~\ref{ccl}.

%
% General Discussion
%
\section{General Discussion}
\label{gen}

We proceed by way of example.
Let us remember that, e.g., the \naive\ 
expression for the two-photon decay $4S \to nS$ reads
(see, e.g., p.~2447 of Ref.~\cite{Fl1984} or p.~4 of Ref.~\cite{ChSu2007})
\begin{align}
\label{naive}
& \Gamma_{4S \to nS} = 
\frac{4 \alpha^2}{27 \pi m^2} \,
\!\!\!\!\!\!
\int\limits_0^{E_{4S} - E_{nS}} 
\!\!\!\!\!\!
{\rm d}\omega \, \omega^3 \, (E_{4S} - E_{nS} - \omega)^3 \,
\left| \sum_{\nu} \right.
\nonumber\\[2ex]
& \left. \left( \!
\frac{\left< nS \! \left| \left| \vec{x} \right| \right| \! \nu P  \right> 
\left< \nu P \! \left| \left| \vec{x} \right| \right| \! 4S \right> }
  {E_{4S} - E_{\nu P}  - \omega + {\rm i} \frac{\Gamma^{(1)}_{\nu P}}{2}} 
% \right.
% \right.
% \nonumber\\[2ex]
%
% \left.
% \left.
\! + \!
\frac{\left< nS \! \left| \left| \vec{x} \right| \right| \! \nu P  \right> 
\left< \nu P \! \left| \left| \vec{x} \right| \right| \! 4S \right> }
  {E_{nS} - E_{\nu P} + \omega + {\rm i} \frac{\Gamma^{(1)}_{\nu P}}{2}} \! 
\right) 
\right|^2
\end{align}
with $n = 1,2,3$. We use natural units ($\hbar = c = \epsilon_0 = 1$). For the 
reduced matrix elements, we use the conventions of 
Refs.~\cite{Ed1957,VaMoKh1988}. In the sequel,
we use the terms ``decay rate'' and ``decay width''
synonymously for $\Gamma$, adopting the general
convention that $\Gamma$ is the decay width of 
a decaying state (Gamow vector, Ref.~\cite{dMGa2002}) with time 
dependence $\exp(-\Gamma \, t)$ for the persistence probability.
The expression $\Gamma_{i \to f}$
denotes the two-photon decay rate (\naive\  expression)
from state $|i \rangle$ to state $|f\rangle$,
whereas $\Gamma^{(1)}_i$ denotes the total
one-photon decay rate of state $| i \rangle$
to all possible final states (we here use the dipole/long-wavelength
approximation exclusively in all calculations,
and we restrict ourselves to a nonrelativistic
hydrogenlike ion in all concrete calculations).
The QED expression for the two-photon decay rate~\cite{Je2007,Je2008,JeSu2008} 
reads
\begin{align}
\label{qed}
\Gamma^{(2)}_{4S \to nS} = & \;
\lim_{\epsilon \to 0}
\frac{4 \alpha^2}{27 \pi m^2} \,
\int\limits_0^{E_{4S} - E_{nS}} 
\!\!\!
{\rm d}\omega \, \omega^3 \, (E_{4S} - E_{nS} - \omega)^3 \,
\nonumber\\[2ex]
& \times \left[ \sum_{\nu} \left(
\frac{\left< nS \! \left| \left| \vec{x} \right| \right| \! \nu P  \right> 
\left< \nu P \! \left| \left| \vec{x} \right| \right| \! 4S \right> }
  {E_{4S} - E_{\nu P}  - \omega + {\rm i} \epsilon } 
\right.
\right.
\nonumber\\[2ex]
& \quad 
\left.
\left. + 
\frac{\left< nS \! \left| \left| \vec{x} \right| \right| \! \nu P  \right> 
\left< \nu P \! \left| \left| \vec{x} \right| \right| \! 4S \right> }
  {E_{nS} - E_{\nu P} + \omega + {\rm i} \epsilon } \! \right) 
\right]^2 \,.
\end{align}
The QED two-photon decay rate is denoted $\Gamma^{(2)}_{i\to f}$ 
in the current work in contrast to the \naive\ expression,
which carries no superscript.

In order to understand the difference of the \naive\ and the QED expression, we
should first observe that the decay width (even the one-photon width) in the
denominator of the \naive\ decay width is parametrically suppressed with
respect to the energy differences of atomic states [$\Gamma^{(1)}_{\nu P} \sim
\alpha (Z\alpha)^4$ vs.~$E_{4S} - E_{\nu P} \sim (Z\alpha)^2$ 
in units of the electron mass],
and we can therefore apply the $(Z\alpha)$-expansion to the 
evaluation of the expression \eqref{naive}. Specifically,
we evaluate \eqref{naive} in the limit 
$\Gamma^{(1)}_{\nu P} \ll E_{4S} - E_{\nu P}$,
which is justified because the decay width $\Gamma^{(1)}_{\nu P}$ is of 
higher order in the $(Z\alpha)$-expansion than the 
energy difference $E_{4S} - E_{\nu P}$. Physically speaking,
the approximation is justified because the lifetime of 
a typical atomic level is long on the time scale of the oscillation
period of radiation emitted in a typical atomic transition.
We can thus expand the \naive{} expression
for the two-photon decay width of highly excited
levels for small decay widths of the virtual states,
keeping the physical value of the decay width at all
stages of the calculation.

The first observation to be made is that physically, in the limit of a small
decay width of the intermediate state $\Gamma^{(1)}_{\nu P} \ll E_{4S} - E_{\nu
P}$, we can
intuitively assume that the entire decay rate of the initial state will be due
to one-photon decay from the initial state
to the intermediate, resonant states
(i.e., to exactly those virtual states with intermediate energy between
the initial and the final state of the two-photon process). Indeed,
once the system has decayed to one of the intermediate state, it is ``stuck
there'' in view of the small decay width (long lifetime) of that intermediate
state. Intuitively, we would thus expect that the \naive\ expression for the
two-photon decay rate might contain, as a lower-order term in the
$(Z\alpha)$-expansion, the one-photon decay rate of the initial state.  Below,
we show by an explicit calculation that this is indeed the case, and that the
\naive\ expression for the two-photon decay rate
simply contains the one-photon
decay rate as a lower-order term. 

This slightly oversimplified statement 
is shown to
hold provided one-photon cascade processes through intermediate,
resonant states are
possible for the chosen initial state under investigation,
and provided we sum over all open
cascade channels. (The first condition is not fulfilled, e.g.,
for the $2S \to 1S$ decay, where no intermediate, resonant
states are available and there are no
lower-order terms to subtract.)

In terms of the $(Z\alpha)$-expansion, the lower-order
term is calculated to be of the order of
order $\alpha (Z\alpha)^4$ as opposed to $\alpha^2 (Z\alpha)^6$
(the latter would be the expected order-of-magnitude for the 
two-photon decay rate).
The remaining term from the \naive\  two-photon decay,
i.e. the expression left over after the subtraction of the 
term of order $\alpha (Z\alpha)^4$,
then is the coherent two-photon decay rate,
and this left-over term is equivalent to the 
prediction from QED theory.

A final word on Eq.~\eqref{naive} is in order.
It has been questioned in the literature (e.g., in Ref.~\cite{ChSu2007}),
whether one should use the 
total or a differential decay rate for the propagator
denominators in this 
case. The most physical and most intuitive prescription is to use the 
total decay rate of every single intermediate state
in order to regularize the quadratic singularity.
Indeed, we find here that a very clear
interpretation of the lower-order terms can be given
provided the total decay width of the
intermediate, virtual states is used as a regulator.
Because the {\rm total} decay rate can be approximated
very well by the {\em one}-photon decay rate for 
all the virtual states, we use the 
total {\em one}-photon decay rate $\Gamma^{(1)}_{\nu P}$
as a regulator in the denominators of Eq.~\eqref{naive}.
One would have to change this prescription
only if the virtual $2S$ state were present as a virtual
states because it is metastable 
and decays primarily via two-photon decay
($\Gamma^{(1)}_{2S} = 0$ in the nonrelativistic 
dipole approximation),
but the $2S$ states is not present as a virtual state in our calculations.

%
% Concrete Calculation
%
\section{Concrete Calculation}
\label{go}

First, we recall the mathematical mechanism
by which the quadratic singularities are regularized
according to Refs.~\cite{Je2007,Je2008,JeSu2008}.
We use Feynman's $\ii\epsilon$ prescription
for the propagator denominators and consider the following
model integral [see Eq.~(16) of Ref.~\cite{Je2008}],
\begin{equation}
\label{expr1}
\lim_{\epsilon \to 0}  {\rm Re}
\int\limits_0^1 \dd\omega \, 
\left( \frac{1}{a - \omega + {\rm i}\epsilon} \right)^2 
= \frac{1}{a(a-1)} \,.
\end{equation}
Here, $\omega \in (0,1)$ is a scaled variable that 
corresponds to a scaled photon energy (scaled relative to the 
entire photon energy interval for the decay process).

The \naive{} regularization 
method (insertion of a decay 
rate) corresponds to a natural quantum mechanical
addition of amplitudes for the two photons to be emitted
(in whichever sequence) before taking the square.
In this case, for vanishing regulator $\epsilon \to 0$
(here, $\epsilon$ corresponds to the regularizing 
decay width), we obtain divergent integrals of the form 
\begin{equation}
\label{expr2}
\int\limits_0^1 \dd\omega \, 
\left| \frac{1}{a - \omega + {\rm i} \epsilon} \right|^2 
\;\; \mathop{=}^{\epsilon \to 0} \;\;
\frac{\pi}{\epsilon} + \frac{1}{a(a-1)} + 
{\cal O}(\epsilon^2) \,,
\end{equation}
for $0 < a < 1$. Combining \eqref{expr1}
and~\eqref{expr2}, we have
\begin{equation}
\label{wow}
\int\limits_0^1 \dd\omega 
\left| \frac{1}{a - \omega + {\rm i}\epsilon} \right|^2 
\nonumber\\[2ex]
\mathop{=}^{\epsilon \to 0} \frac{\pi}{\epsilon} + 
\int\limits_0^1 \dd\omega 
\left( \frac{1}{a - \omega + {\rm i}\epsilon} \right)^2 +
{\cal O}(\epsilon^2) ,
\end{equation}
A natural generalization of this formula,
which will be a key to our further considerations, reads
\begin{align}
\label{key}
& \int\limits_0^1 \dd\omega \, 
\left| \frac{f(\omega)}{a - \omega + {\rm i} \epsilon} \right|^2 
\nonumber\\[2ex]
& = \frac{\pi}{\epsilon} \, f(a) + 
\int\limits_0^1 \dd\omega \, 
\left( \frac{f(\omega)}{a - \omega + {\rm i} \epsilon} \right)^2 +
{\cal O}(\epsilon^2) \,,
\end{align}
with $f(\omega) \in \mathbbm{R}$.
Equation~\eqref{key}~follows from \eqref{expr1} and~\eqref{expr2}
and from the observation that the two integrands differ only 
in a small region about $\omega \approx a$.
Further illustrative remarks regarding 
the model examples are given in Appendix~\ref{further}.

We now investigate the naive expression for the 
total two-photon decay width of the $4S$ state,
which is the sum of the $4S \to 1S$, $4S \to 2S$, and
$4S \to 3S$ channels,
\begin{widetext}
\begin{align}
\label{naive1}
& \Gamma_{4S} = 
\Gamma_{4S \to 1S} +
\Gamma_{4S \to 2S} + 
\Gamma_{4S \to 3S} 
\nonumber\\[2ex]
& = \frac{4 \alpha^2}{27 \pi m^2} \,
\int\limits_0^{E_{4S} - E_{1S}} 
{\rm d}\omega \, \omega^3 \, (E_{4S} - E_{1S} - \omega)^3 \,
\left| \sum_{\nu}
\left( \!
\frac{\left< 1S \! \left| \left| \vec{x} \right| \right| \! \nu P  \right> 
\left< \nu P \! \left| \left| \vec{x} \right| \right| \! 4S \right> }
  {E_{4S} - E_{\nu P}  - \omega + 
  \frac12 \, \ii \, \Gamma^{(1)}_{\nu P} } 
+ \frac{\left< 1S \! \left| \left| \vec{x} \right| \right| \! \nu P  \right> 
\left< \nu P \! \left| \left| \vec{x} \right| \right| \! 4S \right> }
  {E_{1S} - E_{\nu P} + \omega + 
  \frac12 \, \ii \, \Gamma^{(1)}_{\nu P} } \! \right) 
\right|^2
\nonumber\\[2ex]
& \quad + \frac{4 \alpha^2}{27 \pi m^2} \,
\int\limits_0^{E_{4S} - E_{2S}} 
{\rm d}\omega \, \omega^3 \, (E_{4S} - E_{2S} - \omega)^3 \,
\left| \sum_{\nu}
\left( \!
\frac{\left< 2S \! \left| \left| \vec{x} \right| \right| \! \nu P  \right> 
\left< \nu P \! \left| \left| \vec{x} \right| \right| \! 4S \right> }
  {E_{4S} - E_{\nu P}  - \omega + \frac12 \, \ii \, \Gamma^{(1)}_{\nu P}} 
+ \frac{\left< 2S \! \left| \left| \vec{x} \right| \right| \! \nu P  \right> 
\left< \nu P \! \left| \left| \vec{x} \right| \right| \! 4S \right> }
  {E_{2S} - E_{\nu P} + \omega + 
  \frac12 \, \ii \, \Gamma^{(1)}_{\nu P}} \! \right) 
\right|^2
\nonumber\\[2ex]
& \quad + \frac{4 \alpha^2}{27 \pi m^2} \,
\int\limits_0^{E_{4S} - E_{3S}} 
{\rm d}\omega \, \omega^3 \, (E_{4S} - E_{3S} - \omega)^3 \,
\left| \sum_{\nu}
\left( \!
\frac{\left< 3S \! \left| \left| \vec{x} \right| \right| \! \nu P  \right> 
\left< \nu P \! \left| \left| \vec{x} \right| \right| \! 4S \right> }
  {E_{4S} - E_{\nu P}  - \omega + 
  \frac12 \, \ii \, \Gamma^{(1)}_{\nu P}} 
+ \frac{\left< 3S \! \left| \left| \vec{x} \right| \right| \! \nu P  \right> 
\left< \nu P \! \left| \left| \vec{x} \right| \right| \! 4S \right> }
  {E_{3S} - E_{\nu P} + \omega + 
  \frac12 \, \ii \, \Gamma^{(1)}_{\nu P}} \! \right) 
\right|^2 \,.
\end{align}
From the sum(s) over $\nu$, we can now single out those terms 
which generate the double poles in the photon energy integration
in the limit $\Gamma^{(1)}_{\nu P} \to 0$,
and a remainder term ${\mathcal R}$, which contains at most 
simple poles in that limit,
\begin{align}
\label{naive2}
\Gamma_{4S} & = 
\frac{4 \alpha^2}{27 \pi m^2} \,
\int\limits_0^{E_{4S} - E_{1S}} 
{\rm d}\omega \, \omega^3 \, (E_{4S} - E_{1S} - \omega)^3 \,
\left| 
\frac{\left< 1S \! \left| \left| \vec{x} \right| \right| \! 2P  \right> 
\left< 2P \! \left| \left| \vec{x} \right| \right| \! 4S \right> }
  {E_{4S} - E_{2P}  - \omega + \frac12 \, \ii \, \Gamma^{(1)}_{2P}} 
+ \frac{\left< 1S \! \left| \left| \vec{x} \right| \right| \! 2P  \right> 
\left< 2P \! \left| \left| \vec{x} \right| \right| \! 4S \right> }
  {E_{1S} - E_{2P} + \omega + \frac12 \, \ii \, \Gamma^{(1)}_{2P}} 
\right|^2
\nonumber\\[2ex]
& \quad + 
\frac{4 \alpha^2}{27 \pi m^2} \,
\int\limits_0^{E_{4S} - E_{1S}} 
{\rm d}\omega \, \omega^3 \, (E_{4S} - E_{1S} - \omega)^3 \,
\left| 
\frac{\left< 1S \! \left| \left| \vec{x} \right| \right| \! 3P  \right> 
\left< 3P \! \left| \left| \vec{x} \right| \right| \! 4S \right> }
  {E_{4S} - E_{3P}  - \omega + \frac12 \, \ii \, \Gamma^{(1)}_{3P}} 
+ \frac{\left< 1S \! \left| \left| \vec{x} \right| \right| \! 3P  \right> 
\left< 3P \! \left| \left| \vec{x} \right| \right| \! 4S \right> }
  {E_{1S} - E_{3P} + \omega + \frac12 \, \ii \, \Gamma^{(1)}_{3P}} 
\right|^2
\nonumber\\[2ex]
& \quad + 
\frac{4 \alpha^2}{27 \pi m^2} \,
\int\limits_0^{E_{4S} - E_{2S}} 
{\rm d}\omega \, \omega^3 \, (E_{4S} - E_{2S} - \omega)^3 \,
\left| 
\frac{\left< 2S \! \left| \left| \vec{x} \right| \right| \! 3P  \right> 
\left< 3P \! \left| \left| \vec{x} \right| \right| \! 4S \right> }
  {E_{4S} - E_{3P}  - \omega + \frac12 \, \ii \, \Gamma^{(1)}_{3P}} 
+ \frac{\left< 2S \! \left| \left| \vec{x} \right| \right| \! 3P  \right> 
\left< 3P \! \left| \left| \vec{x} \right| \right| \! 4S \right> }
  {E_{2S} - E_{3P} + \omega + \frac12 \, \ii \, \Gamma^{(1)}_{3P}} 
\right|^2 + {\mathcal R} \,.
\end{align}
\end{widetext}
Note that the last term in Eq.~\eqref{naive1}, which corresponds to the 
$4S \to 3S$ decay, does not contain any terms which could possibly
lead to cascade processes, 
and is thus entirely contained in ${\mathcal R}$. 
We now use Eq.~\eqref{key} with the identification 
$\epsilon \equiv \frac12 \, \Gamma^{(1)}_{nP}$ 
with $n = 2,3$ as appropriate for the treatment
of the double poles. For the simple poles, including those 
contained in the remainder term ${\mathcal R}$, we 
only need a principal-value regularization according to the 
model integral
\begin{equation}
\label{pv}
{\rm Re} \, \int_0^1 \dd \omega \, \frac{f(\omega)}{a - \omega + \ii \epsilon} =
{\rm P.V.} \, \int_0^1 \dd \omega \, \frac{f(\omega)}{a - \omega} 
+ {\mathcal O}(\epsilon)\,.
\end{equation}
again with the identification $\epsilon \equiv \frac12 \, \Gamma^{(1)}_{nP}$,
as appropriate. The decay width regulators in the denominators
of Eq.~\eqref{naive1} provide for the imaginary parts.
We can ignore terms of order $\epsilon$ in 
Eq.~\eqref{pv} which would
otherwise lead to corrections to the decay rate of order
$\alpha^3 (Z\alpha)^8$.  Our self-explanatory notation for the 
partial one-photon decay rates is $\Gamma_{i \to f}$ 
for the channel from level 
$| i \rangle$ to $| f \rangle$. Using the formulas
\begin{align}
\Gamma^{(1)}_{nS \to mP} =& \;
\frac43 \, \alpha \, (Z\alpha)^4 \, 
(E_{nS} - E_{nP})^3 \,
\left< nS \left| \left| \vec{x} \right| \right| mP  \right>^2 \,,
\nonumber\\[2ex]
\Gamma^{(1)}_{nP \to mS} =& \;
\frac49 \, \alpha \, (Z\alpha)^4 \, 
(E_{nS} - E_{nP})^3 
\left< nS \left| \left| \vec{x} \right| \right| mP  \right>^2 \,,
\end{align}
together with~\eqref{key} and~\eqref{pv},
we can finally reformulate the \naive\ two-photon decay rate 
$\Gamma_{4S}$ as
%
% Prefactor for the decay rates is 4/3 * 4/9 = 16/27
% Prefactor for the double pole is 4/27 * [2/Gamma] [2 poles] = 16/27.
% OK it checks out.
%
\begin{widetext}
\begin{align}
\label{naive3}
\Gamma_{4S} & = 
\frac{1}{\Gamma^{(1)}_{2P}} \, 
\Gamma^{(1)}_{2P \to 1S} \,  \Gamma^{(1)}_{4S \to 2P}  
+ \frac{1}{\Gamma^{(1)}_{3P}} \, 
\Gamma^{(1)}_{3P \to 2S} \,  \Gamma^{(1)}_{4S \to 3P} 
+ \frac{1}{\Gamma^{(1)}_{3P}} \, 
\Gamma^{(1)}_{3P \to 1S} \,  \Gamma^{(1)}_{4S \to 3P} 
\nonumber\\[2ex]
& \quad + \frac{4 \alpha^2}{27 \pi m^2} \,
\int\limits_0^{E_{4S} - E_{1S}} 
{\rm d}\omega \, \omega^3 \, (E_{4S} - E_{1S} - \omega)^3 \,
\left(
\frac{\left< 1S \! \left| \left| \vec{x} \right| \right| \! 2P  \right> 
\left< 2P \! \left| \left| \vec{x} \right| \right| \! 4S \right> }
  {E_{4S} - E_{2P}  - \omega + \frac12 \, \ii \, \Gamma^{(1)}_{2P}} 
+ \frac{\left< 1S \! \left| \left| \vec{x} \right| \right| \! 2P  \right> 
\left< 2P \! \left| \left| \vec{x} \right| \right| \! 4S \right> }
  {E_{1S} - E_{2P} + \omega + \frac12 \, \ii \, \Gamma^{(1)}_{2P}} 
\right)^2
\nonumber\\[2ex]
& \quad + 
\frac{4 \alpha^2}{27 \pi m^2} \,
\int\limits_0^{E_{4S} - E_{1S}} 
{\rm d}\omega \, \omega^3 \, (E_{4S} - E_{1S} - \omega)^3 \,
\left( 
\frac{\left< 1S \! \left| \left| \vec{x} \right| \right| \! 3P  \right> 
\left< 3P \! \left| \left| \vec{x} \right| \right| \! 4S \right> }
  {E_{4S} - E_{3P}  - \omega + \frac12 \, \ii \, \Gamma^{(1)}_{3P}} 
+ \frac{\left< 1S \! \left| \left| \vec{x} \right| \right| \! 3P  \right> 
\left< 3P \! \left| \left| \vec{x} \right| \right| \! 4S \right> }
  {E_{1S} - E_{3P} + \omega + \frac12 \, \ii \, \Gamma^{(1)}_{3P}} 
\right)^2
\nonumber\\[2ex]
& \quad + 
\frac{4 \alpha^2}{27 \pi m^2} \,
\int\limits_0^{E_{4S} - E_{2S}} 
{\rm d}\omega \, \omega^3 \, (E_{4S} - E_{2S} - \omega)^3 \,
\left( 
\frac{\left< 2S \! \left| \left| \vec{x} \right| \right| \! 3P  \right> 
\left< 3P \! \left| \left| \vec{x} \right| \right| \! 4S \right> }
  {E_{4S} - E_{3P}  - \omega + \frac12 \, \ii \, \Gamma^{(1)}_{3P}} 
+ \frac{\left< 2S \! \left| \left| \vec{x} \right| \right| \! 3P  \right> 
\left< 3P \! \left| \left| \vec{x} \right| \right| \! 4S \right> }
  {E_{2S} - E_{3P} + \omega + \frac12 \, \ii \, \Gamma^{(1)}_{3P}} 
\right)^2 
\nonumber\\[2ex]
& \quad + {\mathcal R} + {\mathcal O}(\alpha^3 (Z\alpha)^8) \,.
\end{align}
We can now drastically simplify those terms in the above expression
which are free from photon energy integrals,
\begin{align}
\label{simp}
& \frac{1}{\Gamma^{(1)}_{2P}}  
\Gamma^{(1)}_{2P \to 1S} \Gamma^{(1)}_{4S \to 2P}  
+ \frac{1}{\Gamma^{(1)}_{3P}}  
\Gamma^{(1)}_{3P \to 2S} \Gamma^{(1)}_{4S \to 3P} 
+ \frac{1}{\Gamma^{(1)}_{3P}} 
\Gamma^{(1)}_{3P \to 1S} \Gamma^{(1)}_{4S \to 3P} 
\nonumber\\[2ex]
& \quad = \Gamma^{(1)}_{4S \to 2P}  
+ \frac{1}{\Gamma^{(1)}_{3P}} \
\left( \Gamma^{(1)}_{3P \to 2S}   
+ \Gamma^{(1)}_{3P \to 1S} \right)   \Gamma^{(1)}_{4S \to 3P} 
= \Gamma^{(1)}_{4S \to 2P} + \Gamma^{(1)}_{4S \to 3P} = \Gamma^{(1)}_{4S} \,,
\end{align}
where we have used $\Gamma^{(1)}_{2P} = \Gamma^{(1)}_{2P \to 1S}$.
This consideration implies that the term $\pi/\epsilon$ in 
Eq.~\eqref{naive}, when identified as 
$\epsilon \equiv \frac12 \, \Gamma^{(1)}_{nP}$
(with $n = 2,3$) and applied in the 
model example in Eq.~\eqref{key}, generates the full one-photon width 
$\Gamma^{(1)}_{4S}$, provided we sum over all open two-photon channels.
Thus, we can write for the total \naive\ expression of the 
decay rate, which is composed of the 
sum of the $4S \to 1S$, $4S \to 2S$, and $4S \to 3S$ decays,
\begin{align}
\label{high}
& \Gamma_{4S} = \Gamma^{(1)}_{4S}
+  \frac{4 \alpha^2}{27 \pi m^2} \,
\int\limits_0^{E_{4S} - E_{1S}} 
{\rm d}\omega \, \omega^3 \, (E_{4S} - E_{1S} - \omega)^3 \,
\left[ \sum_{\nu}
\left( \!
\frac{\left< 1S \! \left| \left| \vec{x} \right| \right| \! \nu P  \right> 
\left< \nu P \! \left| \left| \vec{x} \right| \right| \! 4S \right> }
  {E_{4S} - E_{\nu P}  - \omega + \frac12 \, \ii \, \Gamma^{(1)}_{\nu P}} 
+ \frac{\left< 1S \! \left| \left| \vec{x} \right| \right| \! \nu P  \right> 
\left< \nu P \! \left| \left| \vec{x} \right| \right| \! 4S \right> }
  {E_{1S} - E_{\nu P} + \omega + 
  \frac12 \, \ii \, \Gamma^{(1)}_{\nu P}} \! \right) 
\right]^2
\nonumber\\[2ex]
& \quad + \frac{4 \alpha^2}{27 \pi m^2} \,
\int\limits_0^{E_{4S} - E_{2S}} 
{\rm d}\omega \, \omega^3 \, (E_{4S} - E_{2S} - \omega)^3 \,
\left[ \sum_{\nu}
\left( \!
\frac{\left< 2S \! \left| \left| \vec{x} \right| \right| \! \nu P  \right> 
\left< \nu P \! \left| \left| \vec{x} \right| \right| \! 4S \right> }
  {E_{4S} - E_{\nu P}  - \omega + \frac12 \, \ii \, \Gamma^{(1)}_{\nu P}} 
+ \frac{\left< 2S \! \left| \left| \vec{x} \right| \right| \! \nu P  \right> 
\left< \nu P \! \left| \left| \vec{x} \right| \right| \! 4S \right> }
  {E_{2S} - E_{\nu P} + \omega + 
  \frac12 \, \ii \Gamma^{(1)}_{\nu P}} \! \right) 
\right]^2
\nonumber\\[2ex]
& \quad + \frac{4 \alpha^2}{27 \pi m^2} \,
\int\limits_0^{E_{4S} - E_{3S}} 
{\rm d}\omega \, \omega^3 \, (E_{4S} - E_{3S} - \omega)^3 \,
\left[ \sum_{\nu}
\left( \!
\frac{\left< 3S \! \left| \left| \vec{x} \right| \right| \! \nu P  \right> 
\left< \nu P \! \left| \left| \vec{x} \right| \right| \! 4S \right> }
  {E_{4S} - E_{\nu P}  - \omega + 
  \frac12 \, \ii \, \Gamma^{(1)}_{\nu P}} 
+ \frac{\left< 3S \! \left| \left| \vec{x} \right| \right| \! \nu P  \right> 
\left< \nu P \! \left| \left| \vec{x} \right| \right| \! 4S \right> }
  {E_{3S} - E_{\nu P} + \omega + 
  \frac12 \, \ii \, \Gamma^{(1)}_{\nu P}} \! \right) 
\right]^2 
\nonumber\\[2ex]
& \quad 
+ {\mathcal O}(\alpha^3 (Z\alpha)^8) \; = \;
\Gamma^{(1)}_{4S} + 
\Gamma^{(2)}_{4S \to 1S} +
\Gamma^{(2)}_{4S \to 2S} +
\Gamma^{(2)}_{4S \to 3S} + {\mathcal O}(\alpha^3 (Z\alpha)^8) \; = \;
\Gamma^{(1)}_{4S} + 
\Gamma^{(2)}_{4S} + {\mathcal O}(\alpha^3 (Z\alpha)^8) \,.
\end{align}
\end{widetext}
This equation simply means that the \naively{} regularized
two-photon decay rate $\Gamma_{4S}$ can be written as the sum of three
terms: (i) the total one-photon decay rate $\Gamma^{(1)}_{4S}$ 
(which consitutes the spurious lower-order term),
(ii) the QED two-photon correction $\Gamma^{(2)}_{4S}$ and
(iii) higher-order terms ${\mathcal O}(\alpha^3 (Z\alpha)^8)$.
Note that the second, third and forth term in the 
expression following the first equal sign 
of Eq.~\eqref{high}, with the identification $\epsilon \equiv 
\frac12 \, \Gamma^{(1)}_{\nu P}$, are just the QED expressions for the 
two-photon decay widths corresponding to 
$4S \to 1S$, $4S \to 2S$, and $4S \to 3S$, respectively.
As shown in Refs.~\cite{Je2007,Je2008,JeSu2008},
the two-photon decay rates $\Gamma^{(2)}_{4S \to nS}$
are of order $\alpha^2 (Z\alpha)^6$, as they should be.
We also observe that the QED treatment~\cite{Je2007,Je2008,JeSu2008} thus
eliminates the lower-order terms right from the start.

The generalization to, e.g., an initial $5S$ state is straightforward,
but one has to take into account the $5S \to 4P \to 3D$ cascades.
We will refrain from discussing the generalization further,
as the principle of the calculation should be sufficiently addressed
by the example case of the $4S$ state.

%
% Conclusion
%
\section{Conclusion}
\label{ccl}

The central result of this paper is contained in 
Eq.~\eqref{high}. For the initial 
$4S$ state, it is shown that the coherent two-photon
contribution to the decay rate, as given by a QED treatment, is obtained 
from the \naive\ expression for the two-photon decay width, 
given in Eq.~\eqref{naive}, after the subtraction of lower-order
terms whose sum over all open channels 
is exactly equal to the total one-photon decay width of the initial
state. Note, in particular, that the QED treatment eliminates the 
lower-order terms right from the start~\cite{Je2007,JeSu2008,Je2008}.
This result could be anticipated (see Sec.~\ref{gen}) 
based on a physical consideration: The $(Z\alpha)$-expansion
of the decay rate implies an expansion for small
decay widths of the intermediate, resonant states;
this expansion, in turn, physically implies a long 
lifetime for the intermediate, resonant states
(on the time scale given by the oscillation period 
of radiation emitted in a typical atomic transition).
This means that in a first approximation,
the virtual resonant states ``trap'' the 
electrons after a single photon has been emitted.
This again implies that the lower-order term
contained in the \naive{} expression for the 
two-photon decay rate should be equal to the 
one-photon decay rate of the initial state, provided we sum over
all open channels, and provided all possible
one-photon decays of the initial state can participate in such 
two-photon cascades.
The term left over after the subtraction of the lower-order
terms is the two-photon correction to the decay rate, and it 
is of order $\alpha^2 (Z\alpha)^6$ in units of the 
electron mass, as it should be.

These considerations allow us to give a very clear physical
interpretation of the lower-order terms which are to be expected in the 
treatment of a two-photon decay process with virtual
resonant states: Because the addition of the probability
amplitudes in the \naive\ treatment does not 
lead to a clear separation of the sequential two-photon
emission via the intermediate, resonant states from the 
coherent two-photon contribution to the decay rate,
lower-order terms have to be expected, and in view of the above
arguments, these have to be exactly equal to the one-photon decay rate of the 
initial state.
Let us recall what a treatment of the total decay rate
would imply for the decay width of, say, the $4S$ state,
if we were to add the {\em \naive{}} expression
for the two-photon decay rate to the well-established 
result for the one-photon decay rate~\cite{BeSa1957}
of that state. In this case,
because of the presence of the lower-order term 
in the two-photon decay width,
we would double-count the one-photon decay rate of the $4S$ state
and obtain a result for the total decay width which would be twice as large
as the established value.
This procedure cannot be consistent, and a subtraction
is definitely required, as outlined here. 

Note that spurious lower-order terms are a recurrent theme
in a number of QED calculations. E.g., in the first, non-relativistic treatment 
of the bound-electron self-energy~\cite{Be1947}, the 
result was found to diverge only logarithmically with the energy 
of the virtual photon, but only after a more severe linear divergence 
was identified as a low-energy part of the mass renormalization associated
with the low-photon-energy contribution to the Lamb shift
and subtracted (see also Refs.~\cite{Mo1974a,Mo1974b,JeKe2004aop}).
Similarly, in Ref.~\cite{Ba1951,BaBeFe1953}, the authors had to 
subtract several spurious lower-order terms in order to 
obtain the $\alpha (Z\alpha)^5$ correction to the Lamb shift
from the two-Coulomb-vertex forward scattering amplitude
[see especially Eqs.~(36) and (53) of Ref.~\cite{BaBeFe1953}].
Eventually, these spurious terms could be shown 
to cancel against compensating terms
from the one-Coulomb-vertex term, but if the authors had considered
only the two-vertex contribution to the Lamb shift, then they 
would have had to carry out a subtraction, and this is indeed the 
spirit in which most modern Lamb shift calculations are done
[see also the remarks after Eq.~(32) of Ref.~\cite{Pa1998},
where several lower-order subtraction terms are identified and 
subtracted, albeit in a different
context]. While the result reported here could have been
found immediately after the initial treatment 
due to G\"{o}ppert--Mayer~\cite{GM1931},
concepts developed later in the context of the theory of 
renormalization have inspired us to look for possible 
subtraction terms. Note that the two-photon decay rate
can naturally be interpreted as the imaginary part of the 
two-loop self-energy~\cite{Je2007} and therefore also constitutes
an ``energy shift,'' albeit an ``imaginary one,'' 
or an ``imaginary part thereof.'' The additional
subtraction term found here can thus be interpreted as an 
additional renormalization of the imaginary part of the
bound-state energy which becomes necessary if the \naive{}
method of regularization is used for the resonant intermediate states.

Summarizing the results of our investigations,
we conclude that physically sensible lower-order terms
are obtained from the \naive\ expressions for the two-photon decay rates when
the propagator denominators of the intermediate, resonant states are
regularized by assigning to them the {\em total} decay widths of 
those virtual state. This is in agreement with physical
intuition and reassuring. A final remark on gauge invariance:
For an infinitesimal $\ii \epsilon$
in the propagator denominators~\cite{Je2007,JeSu2008},
all expressions are gauge invariant with
respect to length and velocity gauges, as shown in Ref.~\cite{Je2008}.  For
finite, small decay rates acting as regulators, gauge invariance, strictly
speaking, does not hold. However, the gauge-noninvariance is
shifted to higher-order terms $\sim \alpha^3 (Z\alpha)^8$, which are not
relevant from a phenomenological point of view, and thus, with the 
correct prescription for the subtraction of lower-order terms, the 
result for the two-photon correction to the decay rate
can indeed be obtained from the 
\naive\ expression regularized using the decay widths of the virtual,
resonant, intermediate states.

%
% Acknowledgments
%
\section*{Acknowledgment}

Support from Deutsche Forschungsgemeinschaft (DFG) is 
gratefully acknowledged (Heisenberg program).

\appendix

%
% Further Explanatory Remarks
%
\section{Further Explanatory Remarks}
\label{further}

We would like to illustrate the calculation of the 
two-photon decay rate from highly excited states
by some further remarks.

At the heart of the calculation lies 
the model example in \eqref{expr1} which 
gives the basic regularization of the 
two-photon decay rate by the $\ii \epsilon$ prescription.
Surprisingly, the regulation can also be made plausible,
by a principal-value integration, although this is 
not obvious in the case of a quadratic singularity.
In the related treatment of cascade processes in field theory
[see Ref.~\cite{Ri1972} and Eq.~(6.20) ff.~of Ref.~\cite{BaKoSt1998}]
and of the corresponding double poles,
the principal-value integration leads to the result
[see also Eq.~(32) of Ref.~\cite{Je2008}]
\begin{align}
\label{expr3}
& \int_0^1 \dd\omega \,
\left( {\rm P.V.} \frac{1}{a - \omega} \right)^2 = 
\nonumber\\[2ex]
& = \lim_{\eta \to 0} \int_0^1 \dd\omega 
\left( {\rm P.V.} \frac{1}{a - \omega + \eta } \right) \,
\left( {\rm P.V.} \frac{1}{a - \omega} \right) 
\nonumber\\[2ex]
& = \lim_{\eta \to 0} 
\frac{1}{\eta}
\int_0^1 \dd\omega \,
\left( {\rm P.V.} \frac{1}{a - \omega} -
{\rm P.V.} \frac{1}{a - \omega + \eta} \right) 
\nonumber\\[2ex]
& = \lim_{\eta \to 0} 
\frac{1}{\eta} \,
\left[
\ln\left( \frac{a}{1-a} \right) -
\ln\left( \frac{a + \eta}{1 - a - \eta} \right) 
\right]
\nonumber\\[2ex]
& = \frac{1}{a(a-1)} \,.
\end{align}
The results on the right-hand sides of Eqs.~\eqref{expr1} and~\eqref{expr3} are
identifcal.  The term $\pi/\epsilon$ accounts
for the difference among the two results given in Eqs.~\eqref{expr1}
and~\eqref{expr2}; here this 
term is identified as a spurious lower-order term associated with
one-photon decay to the intermediate resonant state(s).  After its subtraction,
the two results using either the QED prescription~\eqref{expr1} or the \naive{}
prescription~\eqref{expr2}  are in agreement.

\begin{figure}[htb]
\includegraphics[width=0.99\linewidth]{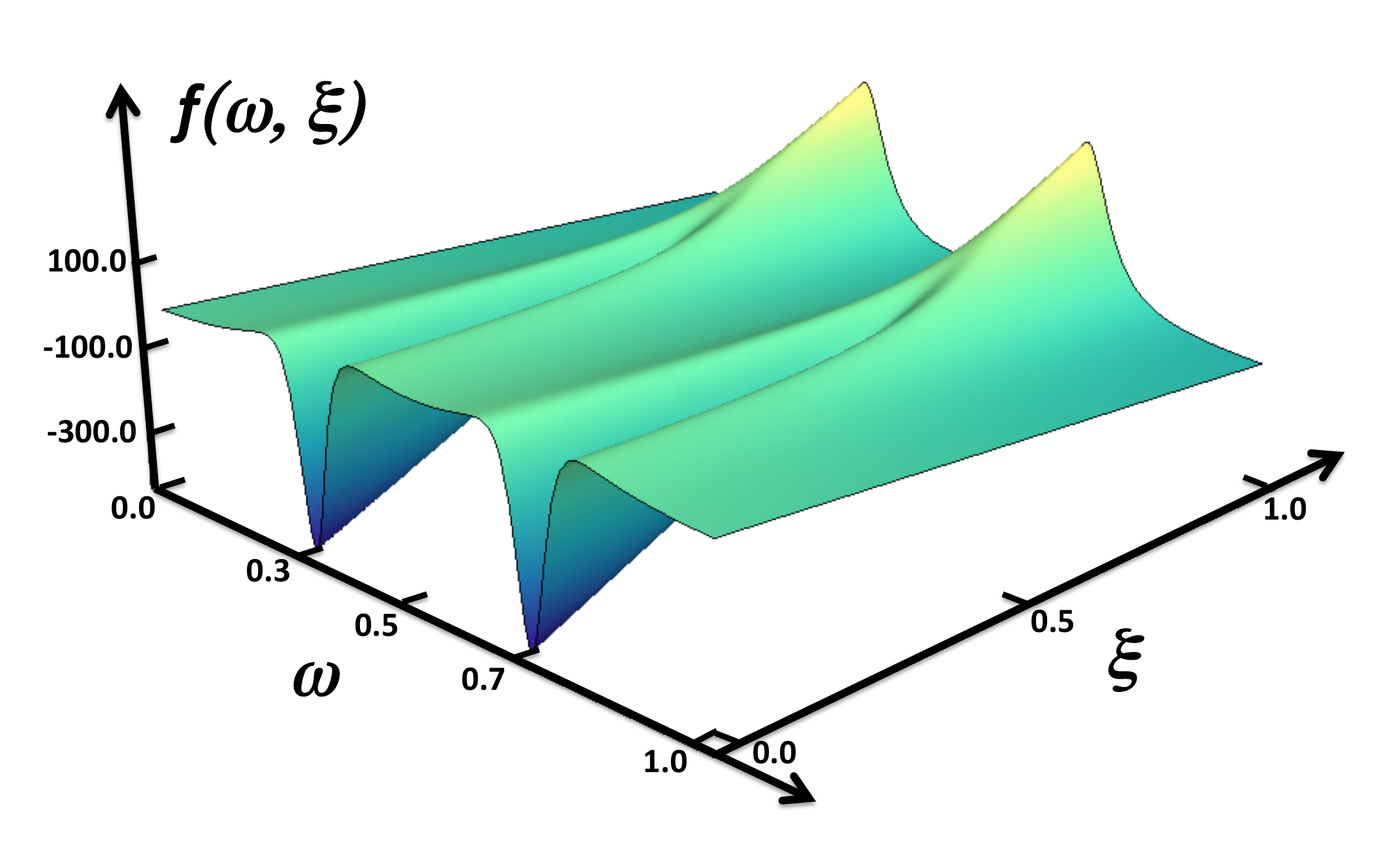} 
\caption{\label{fig1} (Color online) 
Graphical illustration of the subtraction procedure.
The explanation is in the text.}
\end{figure}

The subtraction procedure 
can also be illustrated graphically,
as in Fig.~\ref{fig1}, by plotting a ``model function''
$f(\omega, \xi)$ in terms of its two arguments, which 
we assume to be scaled dimensionless parameters 
$\omega, \xi \in (0,1)$.  For
$\xi = 1$, we plot the \naive{}ly regularized model function 
\begin{equation}
f(\omega, \xi = 1) = 
\left| \frac{1}{\omega_0 - \omega + \ii \epsilon} \right|^2 + 
\left| \frac{1}{\omega_1 - \omega + \ii \epsilon} \right|^2 - 
\frac{ 2 \pi }{ \epsilon } \,,
\end{equation}
illustrating the \naive{}
regularization. This function has two maxima near $\omega_0 =
0.3$ and $\omega_1 = 1 - \omega_0$ which approximate the peaks in the
two-photon spectrum under the presence of a virtual resonant state which is
displaced from the initial state of the two-photon process by the energy
$\omega_0 = 0.3$. We use a numerically small 
regulator value $\epsilon = 0.01$ in accordance with the 
basic assumptions on which our derivation is based;
if $\epsilon$ were large, then, e.g.,
argument regarding the ``trapping'' of the electron in the intermediate
states as given in Sec.~\ref{ccl} would break down. The term
$-2\pi/\epsilon$ is subtracted in order to ensure that the function $f(\omega,
\xi = 1)$ has the same integral over $\omega \in (0,1)$ as the QED regularized
expression 
\begin{equation}
f(\omega, \xi = 0) = {\rm Re} 
\left\{ \frac{1}{(\omega_0 - \omega + \ii \epsilon)^2} + 
        \frac{1}{(\omega_1 - \omega + \ii \epsilon)^2} \right\} \,,
\end{equation}
which has two
minima near $\omega_0$ and $\omega_1$ and illustrates the 
$\ii \epsilon$ prescription. 
The interpolation
\begin{equation}
f(\omega, \xi) = \xi \,\, f(\omega, 1) + (1 - \xi) \,\, f(\omega, 0)
\end{equation}
illustrates that in view of the negative tilt of $f(\omega, \xi)$ along the
$\xi$ axis, the integral $\int_0^1 \dd \omega \, f(\omega, \xi)$, i.e., the
``decay rate,'' is independent of $\xi$ and thus the same in both
regularizations ($\xi = 0$ versus $\xi = 1$).

\end{document}